\documentclass{article}%
\usepackage{amsmath}
\usepackage{amsfonts}
\usepackage{amssymb}
\usepackage{graphicx}%
\providecommand{\U}[1]{\protect\rule{.1in}{.1in}}

 \evensidemargin0in \oddsidemargin0in \topmargin10pt
\begin{document}

\title{Fresnel-transform's quantum correspondence and quantum optical ABCD Law}
\author{Fan Hong-Yi and Hu Li-Yun$^{\ast}$\\{\small Department of Physics, Shanghai Jiao Tong University, Shanghai,
200030, P.R. China}\\$^{\ast}${\small {Corresponding author: hlyun@sjtu.edu.cn.}}}
\maketitle

\begin{abstract}
{\small Corresponding to Fresnel transform there exists a unitary operator in
quantum optics theory, which could be named Fresnel operator (FO). We show
that the multiplication rule of FO naturally leads to the quantum optical ABCD
law. The canonical operator methods as mapping of ray-transfer ABCD matrix is
explicitly shown by FO's normally ordered expansion through the coherent state
representation and the technique of integration within an ordered product of
operators. We show that time evolution of the damping oscillator embodies the
quantum optical ABCD law. }

\end{abstract}


In classical optics, ray-transfer matrices, $M=\left(
\begin{array}
[c]{cc}%
A & B\\
C & D
\end{array}
\right)  ,$ $AD-BC=1$, have been used to describe the geometrical formation of
images by a centered lens system. For an optical ray (a centered spherical
wavefront) passing through optical instruments there is a famous law, named
ABCD law, governing the relation between input ray $\left(  r_{1},\alpha
_{1}\right)  $ and output ray $\left(  r_{2},\alpha_{2}\right)  ,$ i.e.%

\begin{equation}
\left(
\begin{array}
[c]{c}%
r_{2}\\
\alpha_{2}%
\end{array}
\right)  =M\left(
\begin{array}
[c]{c}%
r_{1}\\
\alpha_{1}%
\end{array}
\right)  , \label{1}%
\end{equation}
where $r_{1}$ is the ray height from the optical axis, and $\alpha_{1}$ is
named the optical direction-cosine, $r_{1}/\alpha_{1}\equiv R_{1}$ specifies
the ray's wavefront shape. Eq. (\ref{1}) implies
\begin{equation}
R_{2}\equiv\frac{r_{2}}{\alpha_{2}}=\frac{AR_{1}+B}{CR_{1}+D}. \label{2}%
\end{equation}
This law is the core of matrix optics, since it tells us how the curvature of
a centered spherical wavefront changes from one reference plane to the next.
Besides, the multiplication rule of matrix optics implies that if the
ray-transfer matrices of the $n$ optical components are $M_{1},M_{2}%
,M_{3},\cdots,M_{n}$, respectively, then the whole system is determined by a
matrix $M=M_{1}M_{2}M_{3}\cdots M_{n}.$

One of the remarkable things of modern optics is the case with which
geometrical ray-transfer methods, constituting the matrix optics, can be
adapted to describe the generation and propagation of Laser beams. In 1965
Kogelnik \cite{Kogelnik} pointed out that propagation of Gaussian beam also
obeys ABCD law via optical diffraction integration, i.e. the input light field
$f\left(  x_{1}\right)  $ and output light field $g\left(  x_{2}\right)  $ are
related to each other by so-called Fresnel integration \cite{Goodman}
$g\left(  x_{2}\right)  =\int_{-\infty}^{\infty}\mathcal{K}\left(
A,B,C;x_{2},x_{1}\right)  f\left(  x_{1}\right)  dx_{1},$ where%
\begin{equation}
\mathcal{K}\left(  A,B,C;x_{2},x_{1}\right)  =\frac{1}{\sqrt{2\pi iB}}%
\exp\left[  \frac{i}{2B}\left(  Ax_{1}^{2}-2x_{2}x_{1}+Dx_{2}^{2}\right)
\right]  . \label{3}%
\end{equation}
The ABCD law for Gaussian beam passing through an optical system is
\cite{ABCD}%
\begin{equation}
q_{2}=\frac{Aq_{1}+B}{Cq_{1}+D}, \label{4}%
\end{equation}
where $q_{1}$ $(q_{2})$ represents the complex curvature of the input (output)
Gaussian beam, Eq. (\ref{4}) has the similar form as Eq. (\ref{2}). An
interesting and important question naturally arises: Does ABCD law also
exhibit in quantum optics? Since classical Fresnel transform should have its
quantum optical counterpart (we may name it Fresnel operator (FO)), this
question also challenges us if there exist corresponding multiplication rule
of FO which corresponds to $M=M_{1}M_{2}M_{3}\cdots M_{n}$ ?

In the following we derive ABCD law in quantum optics through introducing the
appropriate FO and exhibiting its multiplication rule. We begin with mapping
the symplectic transform in complex $z$ space $z\rightarrow sz-rz^{\ast}$ onto
operator $U\left(  r,s\right)  $ by virtue of the coherent state
representation \cite{Glauber,Klauder}%
\begin{align}
U\left(  r,s\right)   &  =\sqrt{s}\int\frac{d^{2}z}{\pi}\left\vert
sz-rz^{\ast}\right\rangle \left\langle z\right\vert \nonumber\\
&  \equiv\sqrt{s}\int\frac{d^{2}z}{\pi}\left\vert \left(
\begin{array}
[c]{cc}%
s & -r\\
-r^{\ast} & s^{\ast}%
\end{array}
\right)  \left(
\begin{array}
[c]{c}%
z\\
z^{\ast}%
\end{array}
\right)  \right\rangle \left\langle \left(
\begin{array}
[c]{c}%
z\\
z^{\ast}%
\end{array}
\right)  \right\vert , \label{5}%
\end{align}
where $s$ and $r$ are complex and satisfy the unimodularity condition
$ss^{\ast}-rr^{\ast}=1,$ $\left\vert z\right\rangle =\exp\left(  -\frac{1}%
{2}\left\vert z\right\vert ^{2}+za^{\dagger}\right)  |0\rangle\equiv\left\vert
\left(
\begin{array}
[c]{c}%
z\\
z^{\ast}%
\end{array}
\right)  \right\rangle ,$ $a^{\dagger}$ is the Bose creation operator,
$\left[  a,a^{\dagger}\right]  =1.$ Using the normal ordering of vacuum
projector $\left\vert 0\right\rangle \left\langle 0\right\vert =\colon
\exp\left(  -a^{\dagger}a\right)  \colon\ $and the technique of integration
within an ordered product (IWOP) of operators \cite{IWOP1,IWOP2} we are able
to calculate the integral in Eq.(\ref{5}) and derive its normally ordered
form
\begin{equation}
U\left(  r,s\right)  =\sqrt{\frac{1}{s^{\ast}}}\colon\exp\left[  -\frac
{r}{2s^{\ast}}a^{\dagger2}+\left(  \frac{1}{s^{\ast}}-1\right)  a^{\dagger
}a+\frac{r^{\ast}}{2s^{\ast}}a^{2}\right]  \colon, \label{6}%
\end{equation}
which we name the FO. Using the overlap $\left\langle z\right.  \left\vert
z^{\prime}\right\rangle =\exp[-\frac{1}{2}\left(  \left\vert z\right\vert
^{2}+\left\vert z^{\prime}\right\vert ^{2}\right)  +z^{\ast}z^{\prime}]$ and
the IWOP technique we can obtain multiplication rule of $U\left(  r,s\right)
$,
\begin{equation}
U\left(  r,s\right)  U\left(  r^{\prime},s^{\prime}\right)  =\sqrt{ss^{\prime
}}\int\frac{d^{2}zd^{2}z^{\prime}}{\pi^{2}}\left\vert sz-rz^{\ast
}\right\rangle \left\langle z\right.  \left\vert s^{\prime}z^{\prime
}-r^{\prime}z^{\prime\ast}\right\rangle \left\langle z^{\prime}\right\vert
=U(r^{\prime\prime},s^{\prime\prime}), \label{7}%
\end{equation}
where%
\begin{equation}
\left(
\begin{array}
[c]{cc}%
s & -r\\
-r^{\ast} & s^{\ast}%
\end{array}
\right)  \left(
\begin{array}
[c]{cc}%
s^{\prime} & -r^{\prime}\\
-r^{\prime\ast} & s^{\prime\ast}%
\end{array}
\right)  =\left(
\begin{array}
[c]{cc}%
s^{\prime\prime} & -r^{\prime\prime}\\
-r^{\ast\prime\prime} & s^{\ast\prime\prime}%
\end{array}
\right)  ,\text{ }\left\vert s^{\prime\prime}\right\vert ^{2}-\left\vert
r^{\prime\prime}\right\vert ^{2}=1, \label{8}%
\end{equation}
or $s^{\prime\prime}=ss^{\prime}+rr^{\prime\ast},\;r^{\prime\prime}%
=rs^{\prime\ast}+r^{\prime}s.$ To see the ABCD law more explicitly, we make
the identification $z=\frac{1}{\sqrt{2}}\left(  x+ip\right)  ,$%
\begin{equation}
\left\vert z\right\rangle =\left\vert \left(
\begin{array}
[c]{c}%
x\\
p
\end{array}
\right)  \right\rangle =\exp\left[  i\left(  pX-xP\right)  \right]  \left\vert
0\right\rangle ,\text{ }X=\frac{a+a^{\dagger}}{\sqrt{2}},\text{ }%
P=\frac{a-a^{\dagger}}{\sqrt{2}i}, \label{9}%
\end{equation}
and%
\begin{equation}
s=\frac{1}{2}\left[  A+D-i\left(  B-C\right)  \right]  ,\text{ }r=-\frac{1}%
{2}\left[  A-D+i\left(  B+C\right)  \right]  , \label{10}%
\end{equation}
where the unimodularity $ss^{\ast}-rr^{\ast}=1$ becomes $AD-BC=1,$ which
guarantees the classical Poisson bracket invariant. Accordingly, Eq.(\ref{5})
is re-expressed as%
\begin{align}
U\left(  r,s\right)   &  =\frac{\sqrt{A+D-i(B-C)}}{\sqrt{2}}\int\frac
{dxdp}{2\pi}\left\vert \left(
\begin{array}
[c]{cc}%
A & B\\
C & D
\end{array}
\right)  \left(
\begin{array}
[c]{c}%
x\\
p
\end{array}
\right)  \right\rangle \left\langle \left(
\begin{array}
[c]{c}%
x\\
p
\end{array}
\right)  \right\vert \nonumber\\
&  \equiv F\left(  A,B,C\right)  . \label{11}%
\end{align}
and Eq. (\ref{6}) becomes%
\begin{align}
F\left(  A,B,C\right)   &  =\sqrt{\frac{2}{A+D+i\left(  B-C\right)  }}%
\colon\exp\left\{  \frac{A-D+i\left(  B+C\right)  }{2\left[  A+D+i\left(
B-C\right)  \right]  }a^{\dagger2}\right. \nonumber\\
&  \left.  +\left[  \frac{2}{A+D+i\left(  B-C\right)  }-1\right]  a^{\dagger
}a-\frac{A-D-i\left(  B+C\right)  }{2\left[  A+D+i\left(  B-C\right)  \right]
}a^{2}\right\}  \colon. \label{12}%
\end{align}
From Eq.(\ref{11}) we see that $\left(
\begin{array}
[c]{c}%
x\\
p
\end{array}
\right)  \rightarrow$ $\left(
\begin{array}
[c]{cc}%
A & B\\
C & D
\end{array}
\right)  \left(
\begin{array}
[c]{c}%
x\\
p
\end{array}
\right)  $ in phase space maps onto $F\left(  A,B,C\right)  $. It then follows
from Eqs.(\ref{7}) and (\ref{8}) the multiplication rule for $F$ is $F\left(
A^{\prime},B^{\prime},C^{\prime},D^{\prime}\right)  F\left(  A,B,C,D\right)
=F\left(  A^{\prime\prime},B^{\prime\prime},C^{\prime\prime},D^{\prime\prime
}\right)  ,$ where%
\begin{equation}
\left(
\begin{array}
[c]{cc}%
A^{\prime\prime} & B^{\prime\prime}\\
C^{\prime\prime} & D^{\prime\prime}%
\end{array}
\right)  =\left(
\begin{array}
[c]{cc}%
A^{\prime} & B^{\prime}\\
C^{\prime} & D^{\prime}%
\end{array}
\right)  \left(
\begin{array}
[c]{cc}%
A & B\\
C & D
\end{array}
\right)  =\left(
\begin{array}
[c]{cc}%
A^{\prime}A+B^{\prime}C & A^{\prime}B+B^{\prime}D\\
C^{\prime}A+D^{\prime}C & C^{\prime}B+D^{\prime}D
\end{array}
\right)  . \label{13}%
\end{equation}
To prove that $F$ is just the Fresnel operator responsible for classical
Fresnel transform in Eq.(\ref{3}), we derive $F$'s canonical operator form.
For this aim, we notice that when $B=0$, $A=1,$ $C\rightarrow C/A,$ $D=1,$
Eq.(\ref{12}) becomes%
\begin{equation}
F\left(  1,0,C/A\right)  =\sqrt{\frac{2}{2-iC/A}}\colon\exp\left[  \frac
{iC/A}{2-iC/A}\frac{\left(  a^{\dagger2}+2a^{\dagger}a+a^{2}\right)  }%
{2}\right]  \colon=\exp\left(  \frac{iC}{2A}X^{2}\right)  , \label{14}%
\end{equation}
which is named quadratic phase operator \cite{decom}, where in the last step
we have used the operator identity \cite{IWOP1}
\begin{equation}
e^{\lambda X^{2}}=\frac{1}{\sqrt{1-\lambda}}\colon\exp\left[  \frac{\lambda
}{1-\lambda}X^{2}\right]  \colon. \label{15}%
\end{equation}
When $C=0,$ $A=1,$ $B\rightarrow B/A,$ $D=1,$ Eq.(\ref{12}) reduces to%
\begin{equation}
F\left(  1,B/A,0\right)  =\sqrt{\frac{2}{2+iB/A}}\colon\exp\left[  \frac
{iB/A}{2+iB/A}\frac{\left(  a^{\dagger2}-2a^{\dagger}a+a^{2}\right)  }%
{2}\right]  \colon=\exp\left(  -\frac{iB}{2A}P^{2}\right)  , \label{16}%
\end{equation}
which is named Fresnel propagator in free space, where we have used the
following operator identity
\begin{equation}
e^{\lambda P^{2}}=\frac{1}{\sqrt{1-\lambda}}\colon\exp\left[  \frac{\lambda
}{1-\lambda}P^{2}\right]  \colon. \label{17}%
\end{equation}
When the decomposing is
\begin{equation}
\left(
\begin{array}
[c]{cc}%
A & B\\
C & D
\end{array}
\right)  =\left(
\begin{array}
[c]{cc}%
1 & 0\\
C/A & 1
\end{array}
\right)  \left(
\begin{array}
[c]{cc}%
A & 0\\
0 & A^{-1}%
\end{array}
\right)  \left(
\begin{array}
[c]{cc}%
1 & B/A\\
0 & 1
\end{array}
\right)  , \label{18}%
\end{equation}
we immediately see $F\left(  A,B,C\right)  $ having its canonical operator
$\left(  X,P\right)  $ representation
\begin{align}
F\left(  A,B,C\right)   &  =F\left(  1,0,C/A\right)  F\left(  A,0,0\right)
F\left(  1,B/A,0\right) \nonumber\\
&  =\exp\left(  \frac{iC}{2A}X^{2}\right)  \exp\left(  -\frac{i}{2}\left(
XP+PX\right)  \ln A\right)  \exp\left(  -\frac{iB}{2A}P^{2}\right)  ,\text{ }
\label{19}%
\end{align}
here $F\left(  A,0,0\right)  $ is the squeezing operator
\cite{squeeze,squeeze2,sq3}%
\begin{align}
F\left(  A,0,0\right)   &  =\operatorname{sech}^{1/2}\sigma\colon\exp\left[
\frac{1}{2}a^{\dagger2}\tanh\sigma+\left(  \operatorname{sech}\sigma-1\right)
a^{\dagger}a-\frac{1}{2}a^{2}\tanh\sigma\right]  \colon\nonumber\\
&  =\exp\left(  -\frac{i}{2}\left(  XP+PX\right)  \ln A\right)  ,\text{ \ }
\label{20}%
\end{align}
where $A\equiv e^{\sigma},$ $\frac{A-A^{-1}}{A+A^{-1}}=\tanh\sigma.$ Using the
canonical operator form of $F$ we can deduce its matrix element in the
coordinate states $\left\vert x\right\rangle $ (its conjugate state is
$\left\vert p\right\rangle $)
\begin{align}
\left\langle x^{\prime}\right\vert F\left(  A,B,C\right)  \left\vert
x\right\rangle  &  =e^{\frac{iC}{2A}x^{\prime2}}\left\langle x^{\prime
}\right\vert \exp\left[  -\frac{i}{2}\left(  XP+PX\right)  \ln A\right]
\int_{-\infty}^{\infty}dpe^{-\frac{iB}{2A}p^{2}}\left\langle p\right.
\left\vert x\right\rangle \nonumber\\
&  =e^{\frac{iC}{2A}x^{\prime2}}\left\langle x^{\prime}\right\vert
\int_{-\infty}^{\infty}\frac{dp}{\sqrt{A}}e^{-\frac{iB}{2A}p^{2}}\left\vert
p/A\right\rangle \left\langle p\right.  \left\vert x\right\rangle \nonumber\\
&  =\frac{1}{2\pi}e^{\frac{iC}{2A}x^{\prime2}}\int_{-\infty}^{\infty}\frac
{dp}{\sqrt{A}}e^{-\frac{iB}{2A}p^{2}+ip\left(  x^{\prime}/A-x\right)
}\nonumber\\
&  =\frac{1}{\sqrt{2\pi iB}}\exp\left[  \frac{i}{2B}\left(  Ax^{2}-2x^{\prime
}x+Dx^{\prime2}\right)  \right]  , \label{21}%
\end{align}
which is just the kernel of Fresnel transform in Eq.(\ref{3}), this is why we
name $F\left(  A,B,C\right)  $ the FO (though it can be named a SU(1,1)
generalized squeezing operator either). The above discussions demonstrate how
to transit classical Fresnel transform to FO (and its decomposition of
canonical operators) through the coherent state and the IWOP technique.

Now we directly use the FO to derive ABCD law in quantum optics. From
Eq.(\ref{12}) we see that the FO generates%

\begin{equation}
F\left(  A,B,C\right)  \left\vert 0\right\rangle =\sqrt{\frac{2}{A+iB-i\left(
C+iD\right)  }}\exp\left\{  \frac{A-D+i\left(  B+C\right)  }{2\left[
A+D+i\left(  B-C\right)  \right]  }a^{\dagger2}\right\}  \left\vert
0\right\rangle , \label{22}%
\end{equation}
if we identify
\begin{equation}
\frac{A-D+i\left(  B+C\right)  }{A+D+i\left(  B-C\right)  }=\frac{q_{1}%
-i}{q_{1}+i}, \label{23}%
\end{equation}
then%
\begin{equation}
F\left(  A,B,C\right)  \left\vert 0\right\rangle =\sqrt{-\frac{2/\left(
C+iD\right)  }{q_{1}+i}}\exp\left[  \frac{q_{1}-i}{2\left(  q_{1}+i\right)
}a^{\dagger2}\right]  \left\vert 0\right\rangle , \label{23a}%
\end{equation}
The solution of Eq.(\ref{23}) is
\begin{equation}
q_{1}\equiv-\frac{A+iB}{C+iD}. \label{24}%
\end{equation}
Let $F\left(  A,B,C\right)  \left\vert 0\right\rangle $ expressed by
(\ref{23a}) be an input state for an optical system which is charactristic by
parameters $A^{\prime},B^{\prime},C^{\prime},D^{\prime},$ then the
\textit{quantum optical ABCD law} states that the output state is
\begin{equation}
F\left(  A^{\prime},B^{\prime},C^{\prime}\right)  F\left(  A,B,C\right)
\left\vert 0\right\rangle =\sqrt{\frac{-2/\left(  C^{\prime\prime}%
+iD^{\prime\prime}\right)  }{q_{2}+i}}\exp\left[  \frac{q_{2}-i}{2\left(
q_{2}+i\right)  }a^{\dagger2}\right]  \left\vert 0\right\rangle , \label{24a}%
\end{equation}
which has the similar form as Eq.(\ref{23a})$,$ where $\left(  C^{\prime
\prime},D^{\prime\prime}\right)  $ is determined by Eq.(\ref{13})$,$ and
\begin{equation}
\bar{q}_{2}=\frac{A^{\prime}\bar{q}_{1}+B^{\prime}}{C^{\prime}\bar{q}%
_{1}+D^{\prime}},\text{ \ }\bar{q}_{i}\equiv-q_{i},\text{ \ }\left(
i=1,2\right)  \label{26}%
\end{equation}
which resembles Eq.(\ref{4}).

Proof:

According to the multiplication rule of two FOs and Eqs.(\ref{12})-(\ref{13})
we have%
\begin{align}
&  F\left(  A^{\prime},B^{\prime},C^{\prime}\right)  F\left(  A,B,C\right)
\left\vert 0\right\rangle \nonumber\\
&  =\sqrt{\frac{2}{A^{\prime\prime}+D^{\prime\prime}+i\left(  B^{\prime\prime
}-C^{\prime\prime}\right)  }}\exp\left\{  \frac{A^{\prime\prime}%
-D^{\prime\prime}+i\left(  B^{\prime\prime}+C^{\prime\prime}\right)
}{2\left[  A^{\prime\prime}+D^{\prime\prime}+i\left(  B^{\prime\prime
}-C^{\prime\prime}\right)  \right]  }a^{\dagger2}\right\}  \left\vert
0\right\rangle \nonumber\\
&  =\sqrt{\frac{2}{A^{\prime}\left(  A+iB\right)  +B^{\prime}\left(
C+iD\right)  -iC^{\prime}\left(  A+iB\right)  -iD^{\prime}\left(  C+iD\right)
}}\nonumber\\
&  \times\exp\left\{  \frac{A^{\prime}\left(  A+iB\right)  +B^{\prime}\left(
C+iD\right)  +iC^{\prime}\left(  A+iB\right)  +iD^{\prime}\left(  C+iD\right)
}{2\left[  A^{\prime}\left(  A+iB\right)  +B^{\prime}\left(  C+iD\right)
-iC^{\prime}\left(  A+iB\right)  -iD^{\prime}\left(  C+iD\right)  \right]
}a^{\dagger2}\right\}  \left\vert 0\right\rangle \nonumber\\
&  =\sqrt{\frac{-2/\left(  C+iD\right)  }{A^{\prime}q_{1}-B^{\prime}-i\left(
C^{\prime}q_{1}-D^{\prime}\right)  }}\exp\left\{  \frac{A^{\prime}%
q_{1}-B^{\prime}+i\left(  C^{\prime}q_{1}-D^{\prime}\right)  }{2\left[
A^{\prime}q_{1}-B^{\prime}-i\left(  C^{\prime}q_{1}-D^{\prime}\right)
\right]  }a^{\dagger2}\right\}  \left\vert 0\right\rangle . \label{25}%
\end{align}
Using Eq.(\ref{24}) we see $\frac{2/\left(  C+iD\right)  }{C^{\prime}%
q_{1}-D^{\prime}}=-2/\left(  C^{\prime\prime}+iD^{\prime\prime}\right)  ,$
together using Eq.(\ref{26}) we can reach Eq.(\ref{24a}), thus the law is
proved. Using Eq. (\ref{24}) we can re-express Eq.(\ref{26}) as
\begin{equation}
q_{2}=-\frac{A^{\prime}(A+iB)+B^{\prime}(C+iD)}{C^{\prime}(A+iB)+D^{\prime
}(C+iD)}=-\frac{A^{\prime\prime}+iB^{\prime\prime}}{C^{\prime\prime
}+iD^{\prime\prime}}, \label{28}%
\end{equation}
which is in consistent to Eq.(\ref{24}). Eqs. (\ref{23a})-(\ref{28}) are
therefore self-consistent.

As an application of quantum optical ABCD law, we apply it to tackle the
time-evolution of a time-dependent harmonic oscillator whose Hamiltonian is
\begin{equation}
H=\frac{1}{2}e^{-2\gamma t}P^{2}+\frac{1}{2}\omega_{0}^{2}e^{2\gamma t}%
X^{2},\text{ \ \ }\hbar=1, \label{29}%
\end{equation}
where we have set the initial mass $m_{0}=1,$ $\gamma$ denotes damping. Using
$u\left(  t\right)  =e^{\frac{i\gamma}{2}X^{2}}e^{-\frac{i\gamma t}{2}\left(
XP+PX\right)  }\ $to perform the transformation%
\begin{align}
u\left(  t\right)  Xu^{-1}\left(  t\right)   &  =e^{-\gamma t}X,\nonumber\\
u\left(  t\right)  Pu^{-1}\left(  t\right)   &  =e^{\gamma t}P-\gamma
e^{\gamma t}X, \label{30}%
\end{align}
then $i\frac{\partial\left\vert \psi\left(  t\right)  \right\rangle }{\partial
t}=H\left\vert \psi\left(  t\right)  \right\rangle \ $leads to $i\frac
{\partial\left\vert \phi\right\rangle }{\partial t}=\mathcal{H}\left\vert
\phi\right\rangle ,$ $\left\vert \phi\right\rangle =u\left(  t\right)
\left\vert \psi\left(  t\right)  \right\rangle ,$
\begin{equation}
H\rightarrow\mathcal{H}=u\left(  t\right)  Hu^{-1}\left(  t\right)  -iu\left(
t\right)  \frac{\partial u^{-1}\left(  t\right)  }{\partial t}=\frac{1}%
{2}P^{2}+\frac{1}{2}\omega^{2}X^{2},. \label{31}%
\end{equation}
where $\omega^{2}=\omega_{0}^{2}-\gamma^{2}.$ $\mathcal{H}$ does not contain
$t$ explicitly. The dynamic evolution of a mass-varying harmonic oscillator
from the Fock state $\left\vert 0\right\rangle $ at initial time to a squeezed
state at time $t$ is
\begin{equation}
\left\vert \psi\left(  t\right)  \right\rangle _{0}=u^{-1}\left(  t\right)
\left\vert 0\right\rangle =e^{\frac{i\gamma t}{2}\left(  XP+PX\right)
}e^{-\frac{i\gamma}{2}X^{2}}\left\vert 0\right\rangle , \label{32}%
\end{equation}
if we let $A=D=1,B=0,C=-\gamma;$ and $A^{\prime}=e^{-\gamma t},D^{\prime
}=e^{\gamma t},B^{\prime}=C^{\prime}=0,$ then $q_{1}=\frac{1}{\gamma-i},$
$q_{2}=\frac{e^{-2\gamma t}}{\gamma-i}$, according to Eq.(\ref{24a}) we
directly obtain%
\begin{equation}
u^{-1}\left(  t\right)  \left\vert 0\right\rangle =\sqrt{\frac{2e^{-\gamma t}%
}{e^{-2\gamma t}+i\gamma+1}}\exp\left[  \frac{e^{-2\gamma t}-1-i\gamma
}{2\left(  e^{-2\gamma t}+1+i\gamma\right)  }a^{\dagger2}\right]  \left\vert
0\right\rangle , \label{33}%
\end{equation}
so the time evolution of the damping oscillator embodies the quantum optical
ABCD law.

In summary, the operator methods as mapping of ray-transfer ABCD matrix has
been explicitly shown through FO's normally ordered form and the coherent
state representation. The multiplication rule of FO naturally leads to the
quantum optical version of ABCD law in classical optics. Therefore, the ABCD
law not only exists in classical optics, but also exhibits in quantum optics,
this is a new resemblance between the two fields. The fractional Hankel
transform studied in the context of quantum optics is shown in
Refs.\cite{Hankel,Hankel1}.

This work was supported by the National Natural Science Foundation of China
under grant 10475056.

\end{document}